\documentstyle[aps,prb,twocolumn]{revtex}
\input epsf.sty

\begin{document}
\draft
\title{
Effects of the Nearest-Neighbour Coulomb Interactions on the Ground State of
the Periodic Anderson Model }

\author{ Sushil Lamba, R. Kishore}

\address {Laborat\'orio Associado de Sensores e Materiais, Instituto Nacional de Pesquisas Espaciais,
12 225 Sa\~o Jos\'e dos Campos, S\~ao Paulo, Brazil}
\author{S. K. Joshi}
\address{National Physical Laboratory, Dr.  K.
S. Krishnan Marg, New Delhi 110 012,
India and Jawaharlal
Nehru Center for Advanced Scientific Research,
IISc Campus, Bangalore 560
012, India.}
\date{August 4, 1997}

\maketitle

\begin{abstract}
The magnetic and non-magnetic ground states of the periodic
Anderson model with Coulomb interaction between $f$-electrons on the
nearest-neighbour(NN) sites are investigated using a variational method,
which gives exact calculation of the expectation values in the limit of
infinite dimensions. It is shown that for a critical value of NN Coulomb
interactions the magnetic ground state of the periodic Anderson model in the
Kondo regime is unstable. Factors in terms of the physical processes
responsible for instability of the magnetic ground state are also discussed.
Our study indicates the importance of the NN Coulomb interactions
for correlated two band models.
\end{abstract}


Over the past decade lot of effort has been devoted to the theoretical
understanding of the ground state properties of the heavy-fermion systems.
One of the intriguing experimentally observed phenomena in the heavy-fermion
materials is the variety of magnetic and non-magnetic ground states observed
in these materials~\cite{aeppli94,grewe,ottm,stew,fulde,lee}.  Most of
the theoretical investigations of the magnetic properties
are done on the basis of the periodic Anderson model(PAM) assuming that this
model contains the essential physics of these materials. Theoretical
approaches based on the slave-boson techniques~\cite{read1,read2,coleman84}
are biased towards a paramagnetic ground state while variational
approaches based on the Gutzwiller method  are biased towards a magnetic state. These two
approaches are equivalent in the limit of large orbital degeneracy.
Recently, Reynolds et. al.~\cite{reynolds}, studied the magnetic properties
of the orbitally non-degenerate periodic Anderson model using Kotliar and
Ruckenstein slave-boson(KRSB) formulation of the Gutzwiller method. In
this approach the Gutzwiller approximation is reproduced at the saddle point
for $T=0$. They found that a magnetic instability exists in the entire Kondo
regime and therefore, the Gutzwiller approximation is too biased towards the
magnetic ground state. The
experimental evidence points to the gross inadequacy of the existing
approaches to describe the magnetic behaviour of heavy fermions.

In addition to the
on-site Coulomb interaction in the $f$-band, the other most important
interactions which may affect the stability of the magnetic ground state of the  PAM
are the on-site Coulomb interaction in conduction band and the NN Coulomb
interaction in the $f$-band. The influence of the on-site Coulomb
interaction in the conduction band was recently considered by Itai and Fazekas~\cite{fazekas96} using the
Gutzwiller variational method. They found that this
interaction  reduces the Kondo scale. The reduced
Kondo scale implies that the transitions of electrons from the $f$-band
to the conduction band and vice versa, are further restricted by the presence
of the Coulomb interactions in the conduction band. This  would
lead to further enhancement of
the magnetic ordering of the ground state of the periodic Anderson model.
Consequently, the ground state of the periodic Anderson model with the on-site
Coulomb interaction in the conduction band would be magnetic in the entire
Kondo regime. In the presence of the NN Coulomb interaction
in the $f$-band all the configurations, having electrons on the NN sites
are energetically unfavourable and  the following
physical processes would be operatin (i) $f$-electrons can avoid NN Coulomb
repulsion by occupying next to nearest neighbour sites. This process is
expected to be important only when sufficient number of vacant sites are
available. (ii) Electrons from the $f$-band may go to the Fermi level
whereby take advantage of the hybridization interaction to delocalize.
(iii) Spin-flip process in the $f$-band through hybridization
interaction would also lead to energy gain. All these processes would affect
the magnetic ordering of the ground state of the PAM. The purpose of this
paper is to investigate the influence of the NN Coulomb interaction in the
$f$-band, on the
magnetic instability of the ground state of the PAM in the Kondo regime. To
the best of our knowledge, this is the first study of the influence of the
NN Coulomb interaction on the ground state properties of the PAM.

We consider the extended periodic Anderson model given by

\begin{eqnarray} H & = &{\sum_{k,\sigma}
{{\epsilon_{k}}}d^{\dag }_{k\sigma}
d_{k\sigma}}+{\sum_{i,\sigma} E_f {\mathaccent 94
n}_{fi\sigma}}+V {\sum_{i,\sigma} (d^{\dag }_{i\sigma}
f_{i\sigma}+h.c)} +
\nonumber \\ & & \nonumber
\\ & & \frac{U}{2} \sum_{i,\sigma} n_{fi\sigma} n_{fi-\sigma}+ G
\sum_{\langle i j\rangle \sigma \sigma^{\prime}} n_{fi\sigma}n_{fj\sigma^{\prime}}
\label{eq:mh} \end{eqnarray}
where $n_{fi\sigma }=f_{i\sigma }^{\dagger }f_{i\sigma }$, $i$ and $j$ are
site indices and $k$ are the wave vectors.  The first four terms constitute the standard PAM and the last term in the Hamiltonian corresponds to the Coulomb
interaction between $f$-electrons on the NN sites. $\sum_{\langle ij\rangle }
$ in the last term denotes that the  sum is taken over NN sites only. The total
density of electrons $n=\left( \sum_{i\sigma }n_{fi\sigma }+n_{di\sigma
}\right) /N$, where $N$ is the total number of lattice sites, is taken to be
$1<n<2$, so that there are enough electrons to fill atleast the $f$-levels,
and the $d$-band filling is variable up to half filling.

To study the magnetic ground state we generalize the variational method
previously used to investigate the paramagnetic regime of the PAM~\cite
{lamba1,lamba2,lamba3}. The generalizations are carried out by
distinguishing the up and down spin electrons in the variational
wave function. In the previous treatment for the paramagnetic regime of the
PAM only the lower two spin-degenerate hybridized quasiparticle bands were
considered, however, for the more general case of magnetism, it is required
to take into consideration all four hybridized quasiparticle bands in the
variational wave function.

To investigate magnetic ground state, we choose the variational wave
function as:

\begin{equation}
\label{eq:msiuc}|\psi_{c}\rangle=\prod_i P_i |\psi_{uc}\rangle
\end{equation}

\noindent Where $|\psi _{uc}\rangle ={{\prod}^{\prime}_{k,k^{\prime },\sigma ,\sigma
^{\prime }}}~l_{k\sigma }^{\dagger }u_{k^{\prime }\sigma
^{\prime }}^{\dagger }|0\rangle $ is the uncorrelated wave function. $
\prod^{\prime }$ denotes the product over all occupied states. $u_{k\sigma
}^{\dagger }$ and $l_{k\sigma }^{\dagger }$ create quasiparticles in the
upper and lower hybridized bands respectively. $l_{k\sigma }^{\dagger
}=\alpha _{k\sigma }d_{k\sigma }^{\dagger }-\beta _{k\sigma }f_{k\sigma
}^{\dagger }$ and $u_{k\sigma }^{\dagger }=\alpha _{k\sigma }f_{k\sigma
}^{\dagger }+\beta _{k\sigma }d_{k\sigma }^{\dagger }$. $\alpha _{k\sigma }$
and $\beta _{k\sigma }$ are variational functions, which denote the
probability amplitude for conduction($d$) and $f$-electrons in the
various quasiparticle bands. The quasiparticle creation operators $
l_{k\sigma }^{\dagger }$ and $u_{k\sigma }^{\dagger }$obey fermion
commutation rule if $\alpha _{k\sigma }^2+\beta _{k\sigma }^2=1$.~ The
variational functions $\alpha _{k\sigma }$ and $\beta _{k\sigma }$ differ
from the choice which diagonalizes the Hamiltonian(Eq. 1) in the absence of
Coulomb interactions. Because the Coulomb interactions between the $f$
-electrons can renormalize  the hybridization interaction between $d$ and $
f$-electron and thereby can  also change the probability amplitudes. The
correlation operator, $P_i$, is introduced to suppress those
configurations in the uncorrelated state which are not energetically
favourable in the presence of Coulomb interactions. The correlation operator~\cite{lamba1,lamba2,lamba3} is given by
\begin{equation}
\label{eq:mp}P_i=1+{\sum_\sigma s_\sigma {n}_{fi\sigma }}-\Big [(1-d)+{
\sum_\sigma s_\sigma }\Big]
{n}_{fi\uparrow }{n}_{fi\downarrow }.
\end{equation}

\begin{figure}\epsfxsize=3.0in\epsfbox{
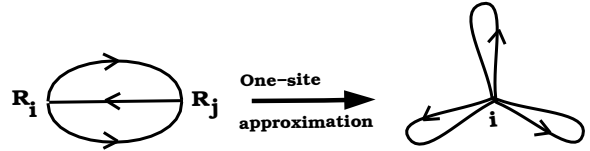
}
\vskip 0.2in

\caption{\em The collapse of intersite diagrams in the one-site
approximation.}
\label{fig:collapse} \end{figure}

The ground state energy per site of the trial wave function(Eq. 2) is given
by $E_g/N={\langle \psi _c|H/N|\psi _c\rangle /{\langle \psi _c|\psi
_c\rangle }}$. The exact calculation of the ground state energy of the
correlated wave function is not possible since the expectation values
involve infinite product of operators and one needs to adopt some
approximate scheme. In this paper we use the  one-site approximation~\cite
{lamba1,lamba2,lamba3} for calculation of various matrix elements appearing
in the ground state energy per site of $|\psi _c\rangle $. The expectation
values appearing in the ground state energy per site of $\psi _c\rangle $,
typically involve expectation values of the type $\langle ...n_{fi\sigma
}n_{fj\sigma ^{\prime }}..\rangle _{uc}$. In the one-site approximation, such expectation values
are approximated by
\begin{equation}
\label{eq:2osa}\langle ...n_{fi\sigma }n_{fj\sigma ^{\prime }}..\rangle
_{uc}=\langle ...\rangle _{uc}~\langle n_{fi\sigma }\rangle _{uc}~\langle
n_{fj\sigma ^{^{\prime }}}\rangle _{uc}~\langle ...\rangle _{uc}
\end{equation}
\noindent where $\langle ....\rangle _{uc}=\langle \psi _{uc}|....|\psi
_{uc}\rangle $. Such an approximation implies the collapse of all
intersite diagrams in the position
space \thinspace [see Figure \ref{fig:collapse}]\thinspace .

\noindent The one-site approximation is expected to give exact calculation
of the expectation values in the limit of infinite dimensions, since as
dimension increases the contribution of the intersite diagrams decreases and
vanishes  altogether in the limit of infinite dimensions~\cite{vol94}.

Using the one-site approximation described above to calculate the expectation values appearing
in the ground state energy of $|\psi _c\rangle $ and minimizing the energy
functional with respect to the variational functions $\alpha _{k\uparrow }$,
$\alpha _{k\downarrow }$, $\beta _{k\uparrow }$, $\beta _{k\downarrow }$ by
imposing the constraint $\alpha _{k\sigma }^2+\beta _{k\sigma }^2=1$, the
minimum of the ground-state energy per site is given by
\begin{eqnarray}
\frac{E_g}{N}&=&\frac{1}{N}\sum_{k\sigma}\left[\xi^{-}_{k\sigma}\langle
l^{\dagger}_{k\sigma} l_{k\sigma}\rangle+
\xi^{+}_{k\sigma}\langle u^{\dagger}_{k\sigma}
u_{k\sigma}\rangle\right]+
\nonumber
\\ & &
\sum_{\sigma}\mu_{\sigma} n_{f\sigma}+U D+G\sum_\sigma (I_\sigma^2+I_\sigma
I_{-\sigma})\label{eq:egosa} \end{eqnarray}
\noindent where $\xi _{k\sigma }^{\pm }$ describes four hybridized
quasiparticle bands
$$
\xi _{k\sigma }^{\pm }=\frac 12{{\left[ {(\epsilon _k+{\tilde E_{f\sigma
})\pm {\left[ {(\epsilon _k-{\tilde E_{f\sigma }})^2+4{\tilde V_\sigma }^2}
\right] }^{\frac 12}}}\right] }}
$$
and $\langle u_{k\sigma }^{\dagger }u_{k\sigma }\rangle _{uc}$ and $\langle
l_{k\sigma }^{\dagger }l_{k\sigma }\rangle _{uc}$ correspond to the average
occupation of the upper ($\xi _{k\sigma }^{+}$) and lower ($\xi _{k\sigma
}^{-}$) quasiparticle bands. $\tilde E_{f\sigma }=E_f-\mu _\sigma $ is the
renormalized $f$-level energy with the $f$-electron self energy $\mu _\sigma
$ given by
\begin{eqnarray} \mu_{\sigma} &=& - \frac{2}{N} \sum_{k \sigma^{\prime}}
{\tilde V_{\sigma^{\prime}}}\frac{\partial{\tilde V_{\sigma^{\prime}}} }{\partial n_{f\sigma}}
\left[  \frac{\langle
u^{\dagger}_{k\sigma^{\prime}}
u_{k\sigma^{\prime}}\rangle_{uc}-\langle
l^{\dagger}_{k\sigma^{\prime}}
l_{k\sigma^{\prime}}\rangle_{uc}}{\sqrt{
{(\epsilon_k-\tilde{E}_{f\sigma^{\prime}})^2+4{\tilde
V_{\sigma^{\prime}}}^2}}}\right ]-\nonumber \\  & & U
\frac{\partial D}{\partial n_{f\sigma}}-G\frac{\partial
\sum_{\sigma}(I_{\sigma}^2 +\,I_{\sigma}I_{-\sigma} )}{\partial
n_{f\sigma}}.
\label{eq:muo}\end{eqnarray}

\noindent
${\tilde V}_{\sigma}= VR_\sigma$ is  the renormalized hybridization
interaction and $R_\sigma$ is the
renormalization factor. $R_\sigma$ and the average
double occupancy, $D$ of the ground state $|\psi_c\rangle$ are given by $$
R_\sigma=\frac{(1-n_f)}{A}\Big[ (1-n_{f-\sigma}) (1+s_\sigma)+d n_{f-\sigma}
(1+s_{-\sigma})\Big]  $$
$$ D={d^2 {(1-n_f)} n_{f\uparrow}
n_{f\downarrow}}/A$$

\noindent  $I_\sigma =n_{f\sigma }\left[ (1+s_\sigma ^2+n_{f-\sigma
}(d^2-(1+s_\sigma )^2)\right] (1-n_f)/A$ with $A={(1-n_f)+(1-d^2)n_{f
\uparrow }n_{f\downarrow }}$. The density of $f$-electrons $n_{f\sigma }$ is
given by
\begin{equation}
\label{eq:nfo}n_{f\sigma }=\frac 1N\sum_k\left[ \beta _{k\sigma }^2\langle
l_{k\sigma }^{\dagger }l_{k\sigma }\rangle _{uc}+\alpha _{k\sigma }^2\langle
u_{k\sigma }^{\dagger }u_{k\sigma }\rangle _{uc}\right]
\end{equation}
\noindent The weight factors $\beta _{k\sigma }^2$ and $\alpha _{k\sigma }^2$
for $f$-electrons with spin $\sigma $ in the lower and upper quasiparticle
bands respectively are given by

\begin{eqnarray}
\alpha_{k\sigma} =\frac{- (\epsilon_k+{\tilde E_{f\sigma})+
\left[(\epsilon_k-{\tilde E_{f\sigma}})^2+4{\tilde V_{\sigma}}^2
\right]^{\frac{1}{2}} }} { \sqrt{2} \left[(\epsilon_k-{\tilde
E_{f\sigma}})^2+4{\tilde V_{\sigma}}^2\right]^{\frac{1}{4}}
}\nonumber
\\ \beta_{k\sigma} =\frac{ (\epsilon_k+{\tilde E_{f\sigma})+
\left[(\epsilon_k-{\tilde E_{f\sigma}})^2+4{\tilde V_{\sigma}}^2
\right]^{\frac{1}{2}} }} { \sqrt{2} \left[(\epsilon_k-{\tilde
E_{f\sigma}})^2+4{\tilde V_{\sigma}}^2\right]^{\frac{1}{4}} }
  \label{eq:4mini-sol} \end{eqnarray}

\noindent The minimization of the ground state energy with respect to $d$ yield the
following implicit equation for $d$.

\begin{eqnarray} U \frac{\partial D}{\partial d} &=&- \frac{2}{N} \sum_{k \,\,\sigma^{\prime}}
{\tilde V_{\sigma^{\prime}}}\frac{\partial{\tilde V_{\sigma^{\prime}}} }{\partial d}
\left [ \frac{\langle
u^{\dagger}_{k\sigma^{\prime}}
u_{k\sigma^{\prime}}\rangle_{uc}-\langle
l^{\dagger}_{k\sigma^{\prime}}
l_{k\sigma^{\prime}}\rangle_{uc}}{\sqrt{
{(\epsilon_k-\tilde{E}_{f\sigma^{\prime}})^2+4{\tilde
V_{\sigma^{\prime}}}^2}}}\right ] \nonumber \\ &&
-G\frac{\partial \sum_{\sigma}(I_{\sigma}^2
+\,I_{\sigma}I_{-\sigma} )}{\partial d}.
\label{eq:do}\end{eqnarray}

At zero temperatures, we can replace
the distribution function for the lower and upper quasiparticle bands by
unit step functions; $\langle l_{k\sigma }^{\dagger }l_{k\sigma }\rangle
_{uc}=\Theta (-\xi _{k\sigma }^{-}+\nu )$ and $\langle u_{k\sigma }^{\dagger
}u_{k\sigma }\rangle _{uc}=\Theta (-\xi _{k\sigma }^{+}+\nu )$. Here $\Theta $
is the unit step function, and $\nu $ is the Fermi level. $\nu $ is
determined by fixing the density of total number of electrons per site,
$n=\sum_\sigma n_\sigma $. At zero temperatures $n_\sigma $ is given
by the following expression
\begin{equation}
\label{eq:no}n_\sigma ={\frac 1N}{\sum_k\left[ \Theta (-\xi _{k\sigma
}^{-}+\nu )+\Theta (-\xi _{k\sigma }^{+}+\nu )\right] }.
\end{equation}

 Before embarking on the numerical calculations it
would be instructive to compare our approach for the periodic Anderson model
with the KRSB reformulation of the Gutzwiller
method~\cite{reynolds}. We note that the ground-state energy
of our variational wave function in the one-site approximation
and the ground-state energy derived from the KRSB approach have different
expressions for the effective hybridization interaction
and the
average double occupancy of the ground state. The average double occupancy
in the one site approximation and the Gutzwiller approximation are given by $D$
and $d_g$(say) respectively. If we scale $D\rightarrow d_g$ in the
expression for the effective hybridization(${\tilde V}_\sigma $) in our
approach, we find that it reduces to the corresponding expression for the
effective hybridization in the KRSB method. This further implies that
the $f$-electron self-energy($\mu _\sigma $), and the average occupation of
the $f$-orbitals in both the approaches also become the same; thereby the
KRSB ground-state energy functional and the one-site ground state
energy functional are the same under the scaling of average double occupancy
of the ground state. Furthermore, since both the approaches search for the
minimum of the ground-state energy in the same physical parameter space, they must
give the same results at the point of minimum. The equivalence of the two
seemingly different variational methods is surprising. To understand this
equivalence, we reanalyze the Gutzwiller variational
wave function. The Gutzwiller wave function has a long history, dating back
to the work of Gutzwiller in 1960's. The Gutzwiller wave function~\cite{gutz,vol84} is given by $|\psi _g\rangle=g^{\hat{D}}|\psi _o\rangle $ . Recently,
Gebhard~\cite{geb90,geb91} showed that it is more convenient to work with
the following form for the Gutzwiller wave function.
\begin{equation}
\label{eq:gsick}|\psi _{gk}\rangle =g^{\hat{K}}|\phi _o\rangle
\end{equation}
\noindent where $|\phi _o\rangle $ is an arbitrary normalized one-particle
product wave function and ${\hat{K}}={\hat{D}}-\sum_{i\sigma }\mu _{i\sigma }n_{fi\sigma }$,
where $\mu _{i\sigma }$ are the explicit functions of $g$ and the local
occupation of $f$-orbitals, $n_{fi\sigma }=\langle \phi _o|n_{fi\sigma
}|\phi _o\rangle $. $|\psi _o\rangle $ and $|\phi _o\rangle $ are connected
by $|\psi _o\rangle =g^{\sum_{i\sigma }\mu _{i\sigma }n_{fi\sigma }}|\phi
_o\rangle $ For the magnetic case the correlator $g^{\hat{K}}$ can be written as $
\prod_iQ_i$ with $Q_i=1+x n_{fi\uparrow }n_{fi\downarrow }-\sum_\sigma
y_\sigma n_{fi\sigma }$, $x$ and $y_\sigma $ are variational
parameters which depend on the average occupation of the $f$-orbitals. With the
redefinition of the parameters $x$ and $y_{\sigma}$ , the correlation operators $Q_i$ and
$P_i$ (Eq. 3) are the same.
Therefore, the Gutzwiller-Gebhard correlator $g^{\hat{K}}$ in the Gutzwiller
approximation  and our correlator $
\prod_iP_i$  in one-site approximation, describe the same physics. It is
interesting to note that the Gutzwiller approximation gives the exact
calculation of the matrix elements in the limit of infinite dimensions and
gives identical results as obtained by the one-site approximation.
Obviously, one-site approximation is much more physically transparent and
operationally simpler than the Gutzwiller approximation.

Although our variational formalism is valid for arbitrary dimension and
dispersion of conduction electron band, for simplicity we assume a
conduction band with a constant density of states $\rho (\epsilon _k)=1/2W$
lying in the energy interval $-W\leq \epsilon _k\leq W$. $2W$ is the
conduction electron band width. We have also taken infinite-$U$ limit, since
at $U=\infty $ the ground state of the PAM is strongly magnetic with maximum
value of total magnetization~\cite{reynolds}. This is an ideal limit to
investigate the instability of the magnetic ground state in the presence of
the nearest neighbour Coulomb interactions. In our formalism this limit is
affected by putting $d=0$ throughout i.e., by projecting out all the doubly
occupied sites. The total magnetization, $
m=\sum_\sigma \sigma n_\sigma $ for different values of nearest neighbour
interaction $G$, the bare hybridization, $V$ and the total electron density,
$n$, is calculated numerically by solving Eq. (\ref{eq:muo}), Eq.
(\ref{eq:nfo}), and Eq. (\ref{eq:no}) self-consistently for $\mu _{\uparrow
}$, $\mu _{\downarrow }$, $n_{f\uparrow }$, $n_{f\downarrow }$, and $\nu $. The numerical solution of the self-consistent
 equations have more than one solutions corresponding to strong magnetism,
 weak magnetism and paramagnetism. The relevant solution is one with the
 lowest ground state energy.
In the numerical calculations we took the conduction electron bandwidth,
$2W=20eV$ and the $f$-level, $E_f=-1.5eV$ below the middle of the conduction
band. We have taken this particular choice of parameter values for reasons of
comparison with earlier work of Reynolds et al~\cite{reynolds} in the absence
of nearest neighbour Coulomb interaction between the $f$-electrons.

In Figure 2,  we have plotted
the total magnetization as a function of the NN Coulomb interaction $G$, for
the total density of electrons, $n$=1.95 and $n$=1.9. For $G/|E_f|=0$, the
ground state is strongly ferromagnetic with total magnetization, $m=0.96$
for $n=1.95$ and $m=.94$ for $n=1.9$. With increasing value of $G$ the
magnetization decreases upto to a critical value of $G$, where we see
a crossover from strong ferromagnetism to weak ferromagnetism with total
magnetization $m=2-n$ and then from weak ferromagnetism to paramagnetism.
Figure 3 shows the magnetic phase diagram of the extended periodic Anderson
model.

In Figure 4 we have plotted the hybridized quasiparticle bands $\xi
_{k\sigma }^{\pm }$ for strongly ferromagnetic ($G/|E_f|=0$), weakly ferromagnetic  ($
G/|E_f|=2$) and paramagnetic ($G/|E_f|=5.6$) ground states. We find that due to the
renormalization of the hybridization interaction and the $f$-electron energy
there is a redistribution of the density of states and to accommodate the
redistribution of density of states the Fermi level also moves to
keep the total density of electrons fixed. For the strongly magnetic ground
state the Fermi level lies in the lower down spin hybridized band($\xi
_{k\downarrow }^{-}$) and the upper up-spin hybridized band($\xi _{k\uparrow
}^{+}$). The lower up-spin hybridized band ($\xi _{k\uparrow }^{-}$) is
completely full.  At a
critical value of $G/|E_f|$ all the electrons in the upper hybridized
up-spin band are transferred to the lower hybridized down-spin band. Then we
see a crossover from strongly ferromagnetic to weakly ferromagnetic ground
state. In the weakly ferromagnetic ground state, the lower hybridized
up-spin band is completely full with total density of up-spin electrons, $
n_{\uparrow }=1$, therefore the total magnetization, $m=n_{\uparrow
}-(n-n_{\uparrow })=2-n$. The total magnetization remains unchanged in the
entire weak ferromagnetic regime till the fermi level also lies in the lower
hybridized up-spin band.
To understand the magnitude
of jump in the magnetization at the point of crossover from strong
ferromagnetism to weak ferromagnetism and then another crossover from weak
ferromagnetism to paramagnetism, it would be instructive to calculate the
density of states $\rho _\sigma ^{\pm }(\omega )$ of hybridized bands, $\xi
_{k\sigma }^{\pm }$. It is given by
\begin{equation}
\rho _\sigma ^{\pm }(\omega )=\sum_k\delta \left(\omega - \xi _{k\sigma }^{\pm
}\right) =\sum_i\delta \left( \omega -x_i\right) \left| \frac{\partial \xi
_{k\sigma }^{\pm }}{\partial \epsilon _k}\right|_{\epsilon _k=x_i}^{-1}
\end{equation}
where $x_i$'s are the roots of $\omega -\xi _{k\sigma }^{\pm }=0$. At a
given energy $\omega $, the density of states is proportional to the slope
of the hybridized bands at $\omega $. The crossover from strong to weak
ferromagnetism is due to shift of the Fermi level from the upper hybridized
up-spin band with larger density of states to the lower hybridized down-spin
band with smaller density of states, resulting in a large decrease in the number of up-spin electrons. In
the crossover from weak ferromagnetism to the paramagnetism,
the change of the
Fermi level in the lower
  hybridized up-spin band to the lower degenerate up and
 down spin bands acompanies a relatively weaker change of the density of
 states.

Figures 5 shows how the NN Coulomb interaction renormalizes the average
occupation of the $f$-orbitals. In the absence of the NN Coulomb interaction
the total magnetization, $m\approx n_{f\uparrow }$. In the strongly
ferromagnetic regime($G/|E_f|<2.0$), for small values of $G$, up-spin
electrons from the $f$-band are transferred to the conduction band through
the hybridization interaction. With increasing values of $G$, more and more
vacant sites are available and spin flip processes  through hybridization
interaction becomes energetically favourable. Therefore, we see an increase
in the number of down-spin electrons and a decrease in the total
magnetization. The weak ferromagnetic regime is stabilized by energy gain
through the transfer of electrons from the $f$-band to the conduction band and by
occupying next to nearest neighbour $f$-electron sites, since there are
sufficiently large number of vacant sites available in this regime.

In this paper we have investigated the magnetic and non-magnetic ground
states of the extended periodic Anderson model, using a variational method based on the
one site approximation. We have shown through the calculation of the magnetic
phase diagram that for $U=\infty$, the non-magnetic ground state is stabilized above a
critical value of nearest neighbour Coulomb repulsion between the
$f$-electrons.  The one site approximation used in this paper gives exact calculation of the
matrix elements in the limit of infinite dimensions. It will be very
interesting to investigate the magnetic properties of the periodic Anderson
model by including  the (dimension)$^{-1}$ contributions through two-site
approximation~\cite{lamba2}. Certainly, it is desirable to extend our
calculations to study antiferromagnetic ground states also.

One of us (S.L) thanks Conselho Nacional de Desenvolvimento Cientifico e
Technologico(CNPq), Brasil for financial assistance.


\begin{figure}\begin{center}
\input{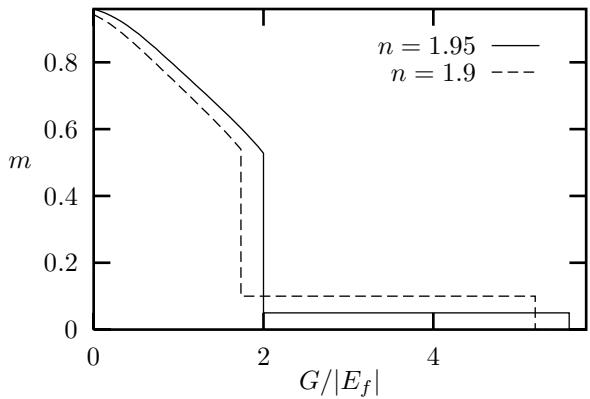}
\end{center}
\caption{\em
{The total magnetization as a function of the Coulomb repulsion between
$f$-electrons on the nearest neighbour sites.
Here we have taken the on-site Coulomb interaction to be
infinitely large and bare hybridization, $V=1.$}}\end{figure}

\begin{figure}
{\input{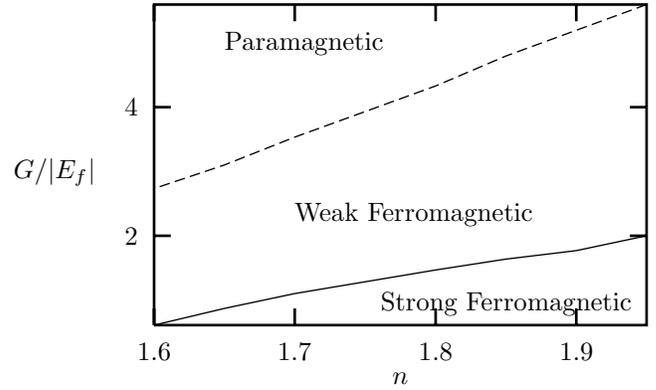}}
\caption{\em
{Magnetic Phase diagram of extended periodic Anderson model}}\end{figure}

\begin{figure}
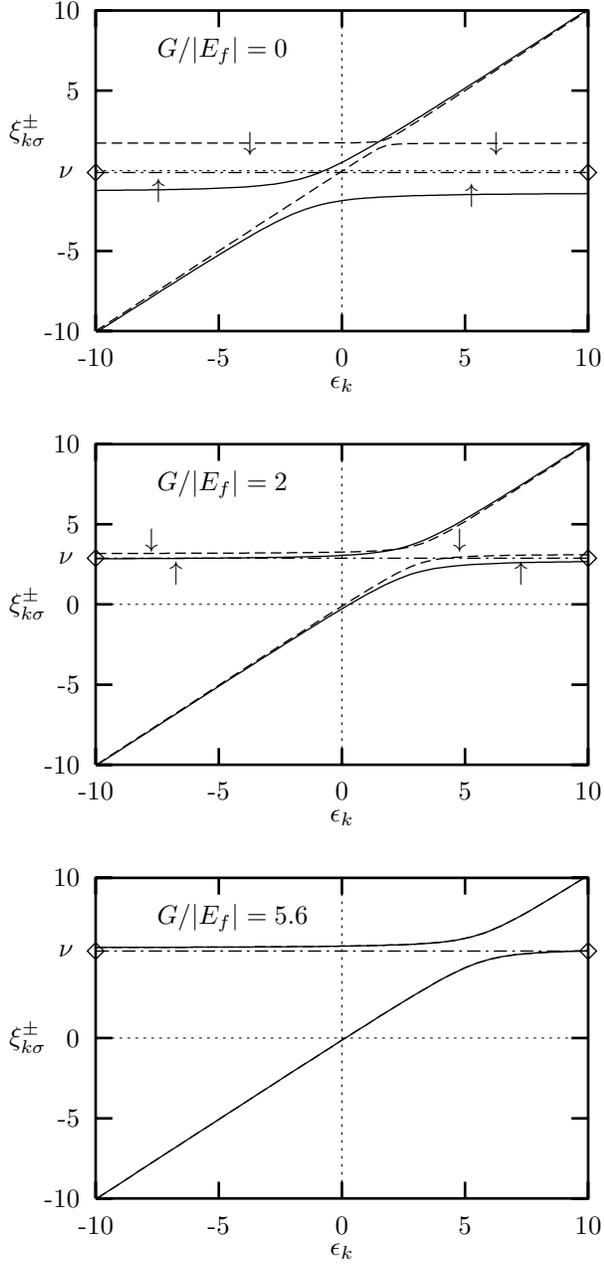
 \begin{center}
\input{fg4a.ps}
\input{fg4b.ps}
\input{fg4c.ps}
\end{center}
\caption{\em
Hybridized quasiparticle bands($\xi^{\pm}_{k\sigma}$) as a
function of conduction band energy($\epsilon_k$) for the total
density of electrons, $n=1.95$ and $V=1.0$. The $\uparrow$ and
$\downarrow$ correspond to the hybridized up-spin band and hybridized
down-spin band respectively. For $G/|E_f|=5.6$, the hybridized band
are spin degenerate.
}  \end{figure}

\begin{figure} \begin{center}
\input{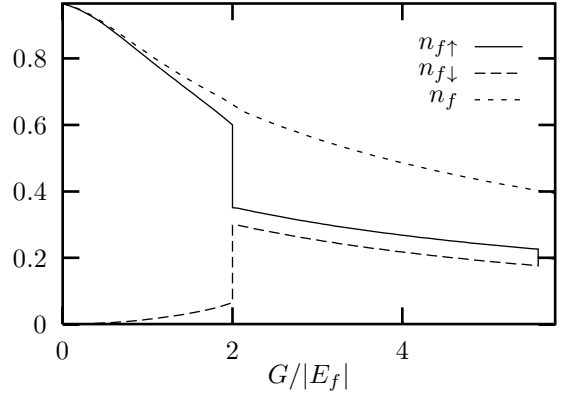}
\end{center}
\caption{\em
The average occupation of the $f$-orbitals $n_{f\sigma}$ are
plotted as a function of $G/|E_f|$ for $n=1.95$.
}  \label{fig:nfs} \end{figure}

\end{document}